\documentclass[12pt]{article}
\usepackage{amsmath}
\usepackage{amsfonts}
\usepackage{epsfig}

\newcommand{\set}[1]{{\mathbb{#1}}}

\begin{document}

\title{Efficient state preparation for a register of quantum bits}
\author{
Andrei N. Soklakov and R\"udiger Schack\\
\\
{\it Department of Mathematics, Royal Holloway,
       University of London,}\\
{\it Egham, Surrey TW20 0EX, United Kingdom}}

\date{11 March 2005}

\maketitle

\begin{abstract}
We describe a quantum algorithm to prepare an arbitrary pure state of a
register of a quantum computer with fidelity arbitrarily close to
1. Our algorithm is based on Grover's quantum search algorithm. 
For sequences of states with suitably bounded amplitudes, the algorithm 
requires resources that are polynomial in the number of qubits.
Such sequences of states occur naturally in the problem of encoding a
classical probability distribution in a quantum register.
\end{abstract}

\section{Introduction}

In many applications of quantum computers, a quantum register, composed of a
fixed number of qubits, is initially prepared in some simple standard state.
This initial preparation step is followed by a sequence of quantum gate
operations and measurements. There are applications of quantum computers, however,
notably the task of simulating the dynamics of a physical system
 \cite{Lloyd1996,Zalka1998,Somaroo1999}, that may require the
initialization of a quantum register in a more general state, corresponding to
the initial physical state of the simulated system. This leads naturally to
the question of what quantum states can be efficiently prepared on a quantum
register.

The memory of a classical computer can be easily put into any state by writing
an arbitrary bit string into it.  The situation for quantum computers is very
different. The Hilbert space associated with a quantum register composed of as
few as 100 quantum bits (qubits) is so large that it is impossible to give a
classical description of a generic state vector, i.e., to list the $2^{100}$
complex coefficients defining it. In this sense it can be said that arbitrary
pure states cannot be prepared \cite{Percival1992}.

It is nevertheless possible to formulate the problem of arbitrary state
preparation for a register of qubits in a meaningful way. This is achieved by
starting from the assumption that the state is initially defined by a set of
quantum {\em oracles}. By assuming that the state is given in this form, we
shift the focus from the problem of describing the state to the problem of the
computational resources needed to actually prepare it. In other words, we
address the computational complexity rather than the algorithmic complexity of
the state. For the purpose of quantifying these computational resources, we
simply count the number of oracle calls. We are thereby deliberately ignoring
the internal structure of the oracles and the computational resources needed 
in each oracle call.
The algorithm we describe here is applicable to any set of oracles, i.e., to
any state. We will show that it is {\em efficient\/} for a large class of
states of particular interest for the simulation of physical systems.

Let $N$ be a positive integer. 
We will describe a quantum algorithm for preparing a $\lceil \log_2
N\rceil$-qubit quantum register in an approximation to the state
\begin{equation}  \label{eq:Psi}
|\Psi\rangle = \sum_{x=0}^{N-1} \sqrt{p(x)} \, e^{2\pi i \phi(x)} |x\rangle
\end{equation}
for arbitrary probabilities $p(x)$ and arbitrary phases $\phi(x)$. Here
and throughout the paper,
$|0\rangle, |1\rangle,\ldots$ denote computational basis states. 
More precisely, given any small positive numbers $\lambda$ and $\nu$, our
algorithm prepares the quantum register in a state $|\tilde\Psi\rangle$
such that, with probability greater than $1-\nu$, the fidelity obeys the bound
\begin{equation}    \label{eq:fidelityBound}
|\langle\tilde\Psi|\Psi\rangle| > 1-\lambda \;.
\end{equation}

To define the algorithm and to assess its efficiency for large $N$, we need to
specify in which form the coefficients $p(x)$ and $\phi(x)$ are given.  We
assume that we are given classical algorithms to compute the functions $p(x)$
and $\phi(x)$ for any $x$. These classical algorithms are used to
construct a set of quantum {\em oracles}. We will  quantify the resources
needed by our state preparation algorithm in terms of (i) the number of oracle
calls, (ii) the number of additional gate operations, and (iii) the number of
auxiliary qubits needed in addition to the $\lceil \log_2 N\rceil$ register
qubits. 

To analyze the asymptotic, large $N$, behavior of our algorithm, we consider a
sequence of probability functions $p_N:\{0,\ldots,N-1\}\to[0,1]$,
$\sum_xp_N(x)=1$, and a sequence of
phase functions $\phi_N:\{0,\ldots,N-1\}\to[0,1]$, where
$N=1,2,\ldots$. For any $N$, the algorithm prepares the quantum register in a
state $|\tilde\Psi\rangle$ such that, with probability greater than $1-\nu$,
the fidelity obeys the bound (\ref{eq:fidelityBound}), where in the
definition~(\ref{eq:Psi}) of $|\Psi\rangle$ the functions $p$ and $\phi$ are
replaced by $p_N$ and $\phi_N$, respectively. Under the assumption that there
exists a real number $\eta$, $0<\eta<1$, such that 
\begin{equation}    \label{eq:etaBound}
p_N(x)\le{1\over\eta N} \;\;\mbox{ for all }\;\; N \;\mbox{ and }\; x \;,
\end{equation}
we show that the resources needed by our
state preparation algorithm are polynomial in the number of qubits, $\log_2N$,
and the inverse parameters $\eta^{-1}$, $\lambda^{-1}$ and $\nu^{-1}$.

An obvious example of a sequence of functions that do not satisfy the
bound~(\ref{eq:etaBound}) and for which the resources required
for state preparation scale exponentially with the number of qubits is given
by $p_N(x)=\delta_{xy}$ for some integer $y=y(N)$. In
this case, it follows {from} the optimality of Grover's algorithm
\cite{Grover1997,Boyer1998} 
that the number of oracle calls needed is proportional to $\sqrt{N}$.

Sequences that do satisfy the bound~(\ref{eq:etaBound}) arise naturally in the
problem of encoding a bounded probability density function
$f:[0,1]\to[0,f_{\rm max}]$ in a state of the form 
\begin{equation}  
|\Psi_f\rangle = {\cal N}^{-1} \sum_{x=0}^{N-1} \sqrt{f(x/N)}\,|x\rangle \;,
\end{equation}
where ${\cal N}$ is a normalization factor. Grover and Rudolph have given an
efficient algorithm for this problem if the function $f$ is efficiently
integrable \cite{Grover-0208}. Essentially the same algorithm was found
independently by Kaye and Mosca \cite{Kaye2001}, who also mention that phase
factors can be introduced using the methods discussed in
Ref.~\cite{Cleve1998}.  Recently, Rudolph \cite{Rudolph-unpublished} has found
a simple nondeterministic state-preparation algorithm that is efficient for
all sequences satisfying the bound~(\ref{eq:etaBound}). 

For general sequences of states satisfying the bound (\ref{eq:etaBound}),
a given value of the fidelity bound $\lambda$,
and assuming polynomial resources for the oracles, our
algorithm is exponentially more efficient than the algorithm proposed by
Ventura and Martinez \cite{Ventura1999} and later related proposals
\cite{Long2001b,Andrecut2001}, for which the resources needed grow like
$N\log_2N$. The use of Grover's algorithm for state preparation has been
suggested by Zeng and Kuang for a special class of coherent states in the
context of ion-trap quantum computers \cite{Zeng2000c}. A general analysis of
the state preparation problem in the context of adiabatic quantum computation
was given by Aharonov and Ta-Shma \cite{Aharonov-0301}.

This paper is organized as follows.  In Sec.~\ref{sec:Algorithm}
we give a full description of our algorithm.  A detailed derivation is
deferred to Sec.~\ref{sec:derivation}.  The algorithm depends on a number of
parameters that can be freely chosen. In Sec.~\ref{sec:fidelity} we
consider a particular choice for these parameters and show that it guarantees
the fidelity bound~(\ref{eq:fidelityBound}).  We use the same choice of
parameters in Sec.~\ref{sec:resources} to derive worst case bounds on the
time and the memory resources required by the algorithm. In
Sec.~\ref{sec:conclusions} we conclude with a brief summary.

\section{Algorithm}\label{sec:Algorithm}

Our algorithm consists of two main stages. 
In the first stage, the algorithm prepares the
register in an approximation to the state 
\begin{equation}    \label{eq:PsiP}
|\Psi_{p}\rangle =
  \sum_{x=0}^{N-1} \sqrt{p(x)} \, |x\rangle \;,
\end{equation}
which differs {from} $|\Psi\rangle$ only in the phases $\phi(x)$. More
precisely, let $\epsilon$ be the largest small parameter such that
\begin{equation}    \label{eq:epsilonBound}
\epsilon<\lambda\eta/3
\end{equation}
and $1/\epsilon$ is an integer.  
The first stage of the algorithm  prepares 
the register in a state
\begin{equation}    \label{eq:PsiPtilde}
|\Psi_{\tilde p}\rangle =
  \sum_{x=0}^{N-1} \sqrt{\tilde p(x)} \, |x\rangle \;,
\end{equation}
such that, with probability greater than $1-\nu$, we 
have that
\begin{equation} \label{TheFidelityBound}
|\langle\Psi_{\tilde p}|\Psi_p\rangle| > 1-\lambda \;.
\end{equation}
We will describe the details of the first stage below.

The second stage of the algorithm adds the phases $\phi(x)$ 
to the state $|\Psi_{\tilde
  p}\rangle$ resulting {from} the first stage. 
This can be done in a straightforward way as follows.
We start by choosing a small parameter $\epsilon'$ such that
$1/\epsilon'$ is a positive integer. We then define a list of unitary
operations, $U_1,\ldots,U_{1/\epsilon'}$, on our quantum register by
\begin{equation} \label{eq:Uk}
U_k|x\rangle=\left\{
           \begin{array}{ll}
e^{2\pi i \epsilon'}|x\rangle     
                 & \mbox{ if}\ \ \phi(x) > (k-{1\over2})\epsilon' \;,\cr
                    |x\rangle & \mbox{ otherwise}.
                   \end{array}
     \right.
\end{equation}
The operators $U_k$ are conditional phase shifts that can be realized as
quantum gate sequences using the classical algorithm for computing the
function $\phi(x)$ \cite{Cleve1998}.  If we apply the operators $U_k$
sequentially to the result of the first stage, we obtain
\begin{equation}
|\tilde\Psi\rangle  =  U_1U_2\cdots U_{1/\epsilon'}|\Psi_{\tilde p}\rangle 
= \sum_{x=0}^{N-1} \sqrt{\tilde p(x)} \, e^{2\pi i \tilde\phi(x)} |x\rangle \;,
\end{equation}
where the function $\tilde\phi(x)$ satisfies the inequality
\begin{equation}   \label{eq:tildePhiBound}
   |\tilde\phi(x) - \phi(x)| \le\epsilon'/2
 \end{equation}
for all $x$. It can be shown (see section~\ref{sec:stage2Fidelity})
that
together with Eq.~(\ref{TheFidelityBound})
this implies the bound
\begin{equation}   \label{eq:properFidelityBound}
|\langle\tilde\Psi|\Psi\rangle| > 1-\lambda-\lambda' \;,
\end{equation}
where $\lambda'={\epsilon'}^2/8$. Notice the slight abuse of notation 
identifying the parameter $\lambda$ in the
inequality~(\ref{eq:fidelityBound}) with the sum $\lambda+\lambda'$ in the
inequality~(\ref{eq:properFidelityBound}). 

\begin{figure}
\epsfig{file=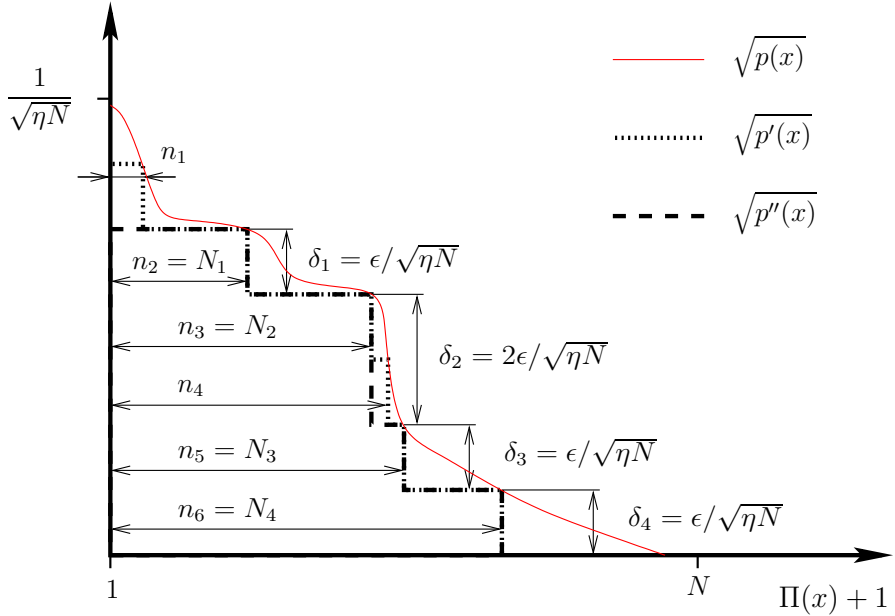,width=12cm}
\caption{(Color online)
  The solid line shows an example for a function $\sqrt{p(x)}$,
  sketched versus a permutation, $\Pi(x)$, chosen so that $p(\Pi^{-1}(x)) \le
  p(\Pi^{-1}(y))$ if $x>y$. The dotted and dashed lines show the corresponding
  functions $\sqrt{p'(x)}$ and $\sqrt{p''(x)}$ [see Eqs.~(\ref{eq:pPrime})
  and~(\ref{eq:pDoublePrime})], which in this representation look like
  decreasing step functions. In this example, $1/\epsilon=7$, $T=4$, and 
  it is assumed that $\tilde
  n_4<\tilde n_3$ due to counting errors. Notice that the function $p(x)$
  is not required to be decreasing.}
\label{figure1}
\end{figure}

We now proceed to a more detailed description of the first stage of the
algorithm. 
{From} now on we assume that $N$ is an integer power of 2. This can always be 
achieved by padding the function $p(x)$ with zeros. 
Given our choice of the parameter $\epsilon$, Eq.~(\ref{eq:epsilonBound}), we define a list of {\em
  oracles}, $o_1,\ldots,o_{1/\epsilon}$, by
\begin{equation} \label{eq:ok}
o_k(x)=\left\{
           \begin{array}{ll}
        1  & \mbox{ if } \sqrt{p(x)} \ge (1-\epsilon k)/\sqrt{\eta N} \;,\cr
        0  & \mbox{ otherwise}.
                   \end{array}
     \right.
\end{equation}
We extend this definition beyond the domain of the function $p$ by setting 
$o_k(x)=0$ for $x\ge N$. 
Using the classical algorithm to compute $p(x)$, one can construct quantum
circuits implementing the unitary oracles 
\begin{equation} \label{eq:okHat}
\hat o_k|x\rangle = (-1)^{o_k(x)}|x\rangle \;.
\end{equation}
These circuits are efficient if the classical algorithm is efficient.
The list of oracles $o_k$ defines a new function $p'(x)$ via
\begin{equation} \label{eq:pPrime}
\sqrt{p'(x)}= (1-\epsilon k)/\sqrt{\eta N} 
\;\;\mbox{if}\;\; 
o_{k-1}(x)=0 \;\;\mbox{and}\;\;
o_{k}(x)=1 \;,
\end{equation}
where $o_0(x)=0$ by convention.
The situation is illustrated in Fig.~\ref{figure1}, where the $x$ values have
been permuted for clarity. Knowledge of this permutation is not required for
our algorithm.

The essence of the first stage of the algorithm consists in using a number of
Grover iterations based on the oracles $\hat o_k$ to prepare the 
register in an approximation to the state
\begin{equation}  \label{eq:PsiPrime}
|\Psi_{p'}\rangle = {\cal N'}^{-1}\sum_{x=0}^{N-1} \sqrt{p'(x)} |x\rangle \;,
\end{equation}
where the normalization factor ${\cal N'}$ reflects the fact that
$p'(x)$ may not be normalized. To find the number of required 
Grover iterations for each oracle $\hat o_k$, 
we need an estimate of the number of
solutions, $n_k$, for each oracle, defined by
\begin{equation}  \label{eq:nk}
n_k = \sum_{x=0}^{N-1} o_k(x) \;.
\end{equation}
This estimate can be obtained {from} running the quantum counting 
algorithm \cite{Boyer1998} for each oracle $\hat o_k$. 
We denote the estimates obtained in this way by $\tilde n_k$.
The accuracy of the estimate $\tilde{n}_k$ relative
to $n_k$ can be characterized by two real parameters,
$0<\nu<1$ and $0<\eta_c<1/2$, in such a way that, as a result of quantum counting,
with probability greater than $1-\nu$ we have
\begin{equation}
|\tilde{n}_k-n_k|<\eta_c N \;\;\mbox{for}\;\;k=1,\ldots,1/\epsilon\;.
\end{equation}
For each oracle, $\hat{o}_k$, the resources needed to 
achieve the counting accuracy specified by $\eta_c$ and $\nu$ 
depend on the actual number of solutions $n_k$.
This dependence is important for optimizing the performance
of our algorithm. In this paper, however, we 
present a simpler analysis
assuming worst case conditions for each oracle $\hat{o}_k$.
For this analysis we use a specific choice of $\eta_c$,
which is given
by Eq.~(\ref{eq:worstCase}).

The analysis of the algorithm is simplified if we concentrate on
a subset of oracles,
\begin{equation}
O_k = o_{f_k} \;\mbox{ for }\; k=1,\ldots,T\;,
\end{equation}
where $T$ and the indices $f_1,\ldots,f_T$ are determined by the construction
below.  We introduce a new parameter $\eta_g$ (see Eq.~(\ref{eq:worstCase}) below) 
such that $\eta_c<\eta_g<1/2$.
The index $f_1$ is defined to  be the smallest integer such that
\begin{equation}    \label{eq:ignorePeak}
\tilde n_{f_1}\ge\eta_gN \;\mbox{ and }\; 
\tilde n_j<\eta_gN \;\mbox{ for }\; j<f_1 \;,
\end{equation}
and, for $k\ge2$, the index $f_k$ is the smallest integer such that 
\begin{equation}    \label{eq:increasing}
 \tilde n_{f_{k-1}} < \tilde n_{f_{k}} \;\mbox{ and }\; 
  f_{k-1} < f_{k} < {1/\epsilon}  \;.
\end{equation}
The number $T$ is the largest value of $k$ for which these inequalities can be
satisfied. 
The effect of Eq.~(\ref{eq:ignorePeak}) is to neglect narrow peaks
(corresponding to small values of $\Pi(x)$ in Fig.~\ref{figure1}).
Equation~(\ref{eq:increasing}) makes sure that the numbers $\tilde n_{f_k}$
form an increasing sequence even if, due to counting errors, the estimates
$\tilde n_k$ do not (see Fig.~\ref{figure1}).

For $k=1,\ldots,T$, define $\tilde{N}_k=\tilde{n}_{f_k}$,
\begin{equation}  \label{eq:defineNk}
N_k=\sum_{x=0}^{N-1}O_k(x)\,,
\end{equation}
and 
\begin{equation} \label{deltas}
\delta_k=\epsilon(f_{k+1}-f_{k})/\sqrt{\eta N}\;,
\end{equation}
where we define $f_{T+1}=1/\epsilon$. 
For every oracle $O_k$, the value of $\tilde N_k$ is an
approximation to the number of solutions, $N_k$, satisfying the bound
\begin{equation}
|\tilde{N}_k-N_k|=|\tilde{n}_{f_k}-n_{f_k}|<\eta_cN\,.
\end{equation}

In what follows it will be convenient to introduce the notation
\begin{equation}   \label{eq:Bsk}
B_{s,k}=\sum_{j=s}^{k}\delta_j\,.
\end{equation}
The oracles $O_k$ define a new function $p''(x)$ via
\begin{equation} \label{eq:pDoublePrime}
\sqrt{p''(x)}= B_{k,T}\;\;
    \mbox{if}\;\; O_{k-1}(x)=0 \;\;\mbox{and}\;\;O_{k}(x)=1\,,
\end{equation}
where $O_0(x)=0$ by convention.
The function $\sqrt{p''(\Pi^{-1}(x))}$ is a decreasing step
function, with step sizes $\delta_1,\ldots,\delta_T$ which are multiples of
$\epsilon/\sqrt{\eta N}$. The widths of the steps are given by the numbers
$N_k$ which are determined by the oracles (see Fig.~\ref{figure1}).

The algorithm can now be completely described as follows. 
Choose a suitable (small) number, $a$, of {\em auxiliary qubits\/} 
(see Eq.~(\ref{eq:worstCase}) below), and define
$L=\log_2 N +a$. For $k=1,\ldots,T$, find the quantities
\begin{equation}   \label{eq:alpha_k}
\tilde\alpha_k^2 =
\frac{\displaystyle\Big(\sum_{s=1}^{k-1}\tilde N_s\delta_s\Big)^2
+\tilde N_k\Big(1-\sum_{s=1}^{k-1}(\tilde N_s-\tilde N_{s-1})B_{s,k-1}^2\Big)
}{\tilde N_k(2^aN-\tilde N_k)}\,,
\end{equation}
\begin{equation}   \label{eq:gamma_fin_and_ini}
\tilde\gamma^{\rm fin}_k=
\frac{\sum_{s=1}^{k}\tilde N_s\delta_s}{\tilde\alpha_k\sqrt{2^aN\tilde{N}_k}}
             \ \ \ {\rm and} \ \ \ 
\tilde\gamma^{\rm ini}_k=
\frac{\sum_{s=1}^{k-1}\tilde N_s\delta_s}
        {\tilde\alpha_k\sqrt{2^aN\tilde{N}_k}}\,,
\end{equation}
\begin{equation}   \label{eq:omega_k}
\tilde{\omega}_k=\arccos\Big(1-\frac{2\tilde{N}_k}{2^aN}\Big)\,,
\end{equation}
and
\begin{equation}    \label{eq:t_k}
t_k = \Big\lfloor {1\over2} + 
{1\over\tilde\omega_k}
(\arcsin\tilde\gamma^{\rm fin}_k - \arcsin\tilde\gamma^{\rm ini}_k )
\Big\rfloor \,.
\end{equation}
For $k=1,\ldots,T$, define the  Grover operator
\begin{equation}
\hat G(O_k,t_k)
  =\left((2|\Psi^0\rangle\langle\Psi^0|-\hat I)\hat O_k\right)^{t_k}\,,
\end{equation}
where  
\begin{equation}  \label{eq:Psi0}
  |\Psi^0\rangle=(2^aN)^{-1/2}\sum_{x=0}^{2^aN-1}|x\rangle \,,
\end{equation}
$\hat I$ is the $L$-qubit identity operator, $\hat
O_k|x\rangle = (-1)^{O_k(x)}|x\rangle$, and where the domain of the oracles
$O_k$ is extended to the range $0\le x<2^aN$ by setting $O_k(x)=0$ if $x\ge
N$.

Prepare a register of $L$ qubits in the state $|\Psi^0\rangle$,
then apply the Grover operators successively to create the state
\begin{equation} \label{eq:psiT}
|\Psi^T\rangle = \hat{G}(O_T,t_T)\cdots\hat{G}(O_1,t_1) |\Psi^0\rangle \;.
\end{equation}
Now measure the $a$ auxiliary qubits in the computational basis.
If all $a$ outcomes are $0$ this stage of the algorithm 
is successfully prepares the desired state Eq.~(\ref{eq:PsiPtilde}).

If one of the measurements of the auxiliary qubits 
returns 1, this stage of the algorithm has failed, and one has to start
again by preparing the register in the state $|\Psi^0\rangle$ as in
Eq.~(\ref{eq:Psi0}). Assuming the choice of parameters in
Eq.~(\ref{eq:worstCase}), the probability, $p_{\rm fail}$, that the
algorithm fails in this way satisfies the bound 
$p_{\rm fail}<10\lambda$ (see Eq.~(\ref{TheFailureProbability})). 

Before we provide detailed proofs of the above claims
it is helpful to give a hint of how this stage of the
algorithm achieves its goal. The algorithm aims
at constructing the function $\sqrt{\tilde p(x)}$ 
which is close to the function $\sqrt{p''(x)}$ 
defined in Eq.~(\ref{eq:pDoublePrime}).  
The sequence of Grover operators in
Eq.~(\ref{eq:psiT}) creates a step function
that is close to $\sqrt{p''(\Pi^{-1}(x))}$ (see Fig.~\ref{figure1}).
In particular each operator $\hat{G}(O_k,t_k)$ in
Eq.~(\ref{eq:psiT}) creates a step with the
correct width $N_k$ and a height $h_k$
which is close to the target height $\delta_k$.
Due to a remarkable property of Grover's 
algorithm~\cite{Biham1999}, once the features
$h_1,\ldots,h_{k-1}$ have been developed
they are not distorted by 
$\hat{G}(O_k,t_k)$ which develops $h_k$.
In this way the algorithm proceeds building
feature after feature until it constructs all
$T$ of them. At the end, because of the inherent
errors, the auxiliary qubits end up having small amplitudes
for nonzero values. Measuring them 
projects the auxiliary qubits onto the zero values
with a probability that can be made arbitrarily close to 1. 
This also slightly changes the features $h_k$
due to renormalization of the state after the 
measurement, which we take into account when we estimate
the overall loss of fidelity.

\section{Derivation of the algorithm}\label{sec:derivation}

In section~\ref{sec:Algorithm} we have already
explained the second stage of our algorithm.
Here we present a detailed explanation
of the first stage. This section is organized
as follows. In subsection~\ref{subsecBiham}
we review some properties
of the Grover operator introducing
our notation as we go along.
In subsection~\ref{subsecOneStep}
we introduce a convenient mathematical
form for analyzing intermediate quantum 
states visited by the algorithm.  
And finally, in subsection~\ref{subsec:t_and_h}
we derive the values~(\ref{eq:t_k}) of the times $t_k$ used in our
algorithm.

\subsection{Preservation of features by the Grover operator}\label{subsecBiham}

  We will be using the following result~\cite{Biham1999}. 
  Consider an oracle $O$, which accepts $r$ values of $x$ (out
  of the total of $2^aN$, i.e., $\sum_{x=0}^{2^aN-1}O(x)=r$). We shall call
  such values of $x$ {\em good}, as opposed to {\em bad\/} values of $x$ that
  are rejected by the oracle. Using different notation for the coefficients of
  good and bad states, we have that after $t$ Grover iterations an arbitrary
  quantum state
\begin{equation} \label{initial}
|\Psi^{\rm ini}\rangle= \sum_{{\rm good\ }x} g^{\rm ini}_x|x\rangle
                           +\sum_{{\rm bad\ }x} b^{\rm ini}_x|x\rangle
\end{equation}
is transformed into
\begin{equation}
|\Psi^{\rm fin}\rangle=\hat{G}(O,t)|\Psi^{\rm ini}\rangle
=\sum_{{\rm good\ }x} g^{\rm fin}_x|x\rangle
                           +\sum_{{\rm bad\ }x} b^{\rm fin}_x|x\rangle.
\end{equation}
Let $\bar{g}^{\rm ini}$ and $\bar{b}^{\rm ini}$ be the averages of the initial
amplitudes of the good and the bad states respectively:
\begin{equation}
\bar{g}^{\rm ini}
=\frac{1}{r}\sum_{{\rm good\ }x}g^{\rm ini}_x\,,\hspace*{2cm}
\bar{b}^{\rm ini}=\frac{1}{2^aN-r}\sum_{{\rm bad\ }x}b^{\rm ini}_x\,,
\end{equation}
and similarly for the final amplitudes
\begin{equation}
\bar{g}^{\rm fin}
=\frac{1}{r}\sum_{{\rm good\ }x}g^{\rm fin}_x\,,\hspace*{2cm}
\bar{b}^{\rm fin}=\frac{1}{2^aN-r}\sum_{{\rm bad\ }x}b^{\rm fin}_x\,,
\end{equation}
Let us also define
\begin{equation}
\Delta g^{\rm ini}_x= 
g^{\rm ini}_x-\bar{g}^{\rm ini}\,,\hspace*{2cm}
\Delta b^{\rm ini}_x=b^{\rm ini}_x-\bar{b}^{\rm ini}\,.
\end{equation}
In other words, $\Delta g^{\rm ini}_x$ and $\Delta b^{\rm ini}_x$ define the
{\em features\/} of the initial amplitude functions $g^{\rm ini}_x$ and
$b^{\rm ini}_x$ relative to their averages $\bar{g}^{\rm ini}$ and
$\bar{b}^{\rm ini}$.  Biham {\it et.\ al.} have shown that the change of the
amplitudes is essentially determined by the change of the averages:
\begin{eqnarray} \label{BihamEquations}
g^{\rm fin}_x&=&\bar{g}^{\rm fin}+\Delta g^{\rm ini}_x \cr
b^{\rm fin}_x&=&\bar{b}^{\rm fin}+(-1)^t\Delta b^{\rm ini}_x\,,
\end{eqnarray}
where the averages $\bar{g}^{\rm fin}$ 
and $\bar{b}^{\rm fin}$ are given as follows. 
Define
\begin{eqnarray}
\omega&=&\arccos\left(1-\frac{2r}{2^{a}N}\right)\;,\cr
\alpha
&=&\sqrt{|\bar{b}^{\rm ini}|^2+|\bar{g}^{\rm ini}|^2\frac{r}{2^aN-r}}\;,\cr
\phi
&=&\arctan\left(\frac{\bar{g}^{\rm ini}}{\bar{b}^{\rm ini}}
    \sqrt{\frac{r}{2^aN-r}}\,\right)\;.
\end{eqnarray} 
The averages are given by
\begin{eqnarray}
\bar{g}^{\rm fin}&=&\sqrt{\frac{2^aN-r}{r}}\;\alpha\,\sin(\omega t+\phi)\,, \cr
\bar{b}^{\rm fin}&=&\alpha\,\cos(\omega t+\phi)\,.
\end{eqnarray}
We shall also use the separations, $\Delta^{\rm ini}$ and
$\Delta^{\rm fin}$, of the averages
\begin{equation}
\Delta^{\rm ini}=\bar{g}^{\rm ini}-\bar{b}^{\rm ini}\,\hspace*{2cm}
\Delta^{\rm fin}=\bar{g}^{\rm fin}-\bar{b}^{\rm fin}\,.
\end{equation}  
Directly {from} the definition we obtain
\begin{equation} \label{DeltaT}
\Delta^{\rm fin}=\frac{\alpha}{\sqrt{r/(2^aN)}}\sin(\omega t-\xi),
\end{equation}
where the phase $\xi$ can be found {from} $\Delta^{\rm ini}$,
\begin{equation}
\xi=-\arcsin\Big{(}
    \frac{\Delta^{\rm ini}}{\alpha}\sqrt{\frac{r}{2^aN}}
        \Big{)}\,.
\end{equation}
In our applications we will only need the case when the initial amplitude of
the bad states is flat, i.e.\ $\Delta b^{\rm ini}_x=0$. In this case, the
amplitude of the bad states always remains flat
\begin{equation}
b_x^{\rm fin}=\bar{b}^{\rm fin}\,,
\end{equation}

This fact and the fact that the initial features, $\Delta g_x^{\rm ini}$, of
the amplitude of the good states are preserved,
\begin{equation}
g_x^{\rm fin}-\bar{g}^{\rm fin}=\Delta g_x^{\rm ini}\,,
\end{equation}
see~(\ref{BihamEquations}), is crucial for
understanding the rest of this paper.

\begin{figure}[here]
\epsfig{file=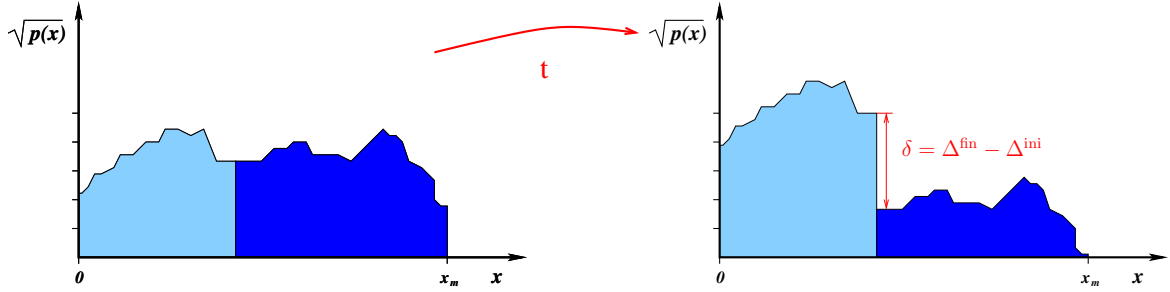,width=15.5cm}
\caption{(Color online)
  Illustration of the action of the Grover operator $\hat G(O,t)$  on an
  arbitrary amplitude function $\sqrt{p(x)}$. The light area indicates values
  of $x$ that are accepted by $O$ (``good values''), and the dark area
  indicates values of $x$ that are  rejected by $O$ (``bad values'').}
\label{BihamFigure}
\end{figure}

\subsection{Intermediate states}\label{subsecOneStep}

The preparation stage of our algorithm summarized in Eq.~(\ref{eq:psiT}) 
gives rise to a sequence of states defined by
\begin{equation} \label{eq:psiK}
|\Psi^k\rangle = \hat{G}(O_k,t_k) |\Psi^{k-1}\rangle \;.
\end{equation}
Let $\set{O}_k$ be the set of solutions to the oracle $O_k$:
\begin{equation}
\set{O}_k=\{x:O_k(x)=1\}\,.
\end{equation}
For $k=2,\dots,T$, these states can be written in the form
\begin{equation}\label{PsiKminusOne}
|\Psi^{k}\rangle=
A_1^{k}\sum_{x\in\set{O}_1}|x\rangle
+A_2^{k}\sum_{x\in\set{O}_2}|x\rangle
+\dots+
A_{k}^{k}\sum_{x\in\set{O}_k}|x\rangle
+B^{k}\sum_{x\not\in\set{O}_k}|x\rangle\,,
\end{equation} 
where
\begin{equation}\label{A_jk}
A_j^{k}=B^{k}+\sum_{s=j}^{k}h_s\,.
\end{equation}
Since $B^{k}$ is real and positive, the value of $B^{k}$ can be determined
{from} the normalization condition $\langle\Psi^{k}|\Psi^{k}\rangle=1$.

\begin{figure}[here]
\epsfig{file=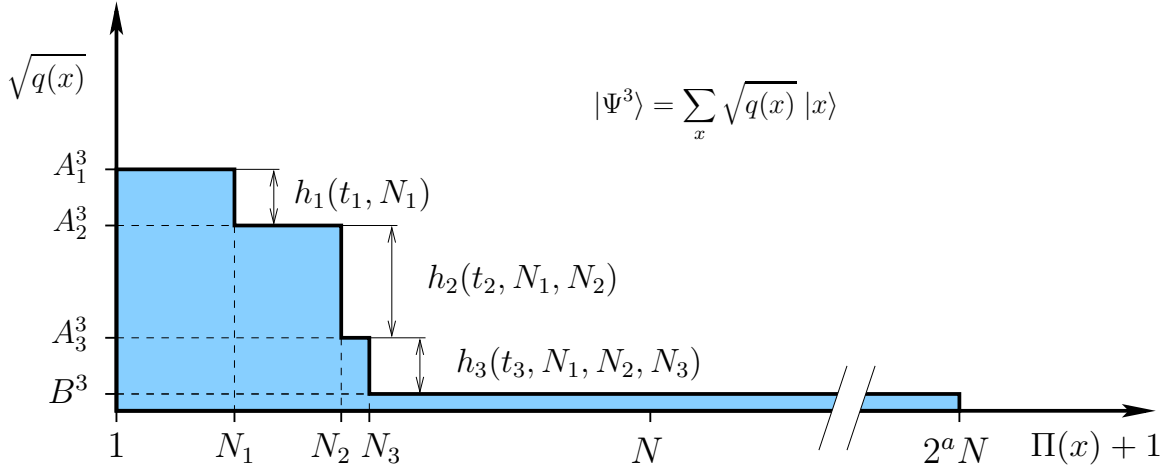,width=15.5cm}
\caption{(Color online)
This figure shows the amplitudes for the state, $|\Psi^{3}\rangle$,
after three iterations of the algorithm. See the text for details.}
\label{UnfinishedStatePrep}
\end{figure}

The action of the algorithm can be visualized as shown in
Fig.~\ref{UnfinishedStatePrep}, which shows the result of the first three
iterations. The integers $N_k$ [defined in Eq.~(\ref{eq:defineNk})] are the
number of good values of $x$ according to oracle $O_k$.  We see that each
operation $\hat{G}(O_k,t_k)$ prepares a feature of height
$h_k=h_{k}(t_k,N_1,\dots,N_k)$.  It follows {from} the conclusions of
section~\ref{subsecBiham} that once such a feature is developed, its height
remains constant throughout the computation.

\subsection{Values of $t_k$} \label{subsec:t_and_h}

In this subsection we show how the times $t_k$ are related to the
corresponding features $h_k$.

The normalization condition $\langle\Psi^{k}|\Psi^{k}\rangle=1$ reads
\begin{equation}
N_1(A_1^{k})^2+(N_2-N_1)(A_2^{k})^2
+\dots+(N_{k}-N_{k-1})(A_{k}^{k})^2
+(2^aN-N_{k})(B^{k})^2=1.
\end{equation}
Substituting (\ref{A_jk}), this gives us a quadratic
equation for $B^{k}$:
\begin{equation}
2^aN(B^{k})^2+2\Big(\sum_{s=1}^{k}N_s h_s\Big)B^{k}
+\sum_{s=1}^{k}(N_s-N_{s-1})C^2_{s,k}-1=0,
\end{equation}
where we define $N_0=0$ and
\begin{equation}
C_{s,k}=\sum_{j=s}^{k}h_j\,.
\end{equation}
Solving this equation, and using the fact that $B^{k}\geq 0$, we obtain
\begin{equation}
B^{k}=-\frac{\sum_{s=1}^{k}N_s h_s}{2^aN}
+\sqrt{\Big(\frac{\sum_{s=1}^{k}N_s h_s}{2^aN}\Big)^2
+\frac{1-\sum_{s=1}^{k}(N_s-N_{s-1})C_{s,k}^2}{2^aN}} \;.
\end{equation}
This formula together with Eq.~(\ref{A_jk}) provide an explicit expression
(\ref{PsiKminusOne}) for $|\Psi^{k}\rangle$ in terms of the numbers
$\{N_s\}_{s=1}^{k}$ and $\{h_s\}_{s=1}^{k}$.

To build the $k$th feature, we apply the Grover operator $\hat{G}(O_k,t_k)$ to
the state $|\Psi^{k-1}\rangle$.  We will now derive an expression for the
integer ``time'' $t_k$ in terms of the features $h_1,\ldots,h_k$ and the
widths $N_1,\ldots,N_k$.  Given $|\Psi^{k-1}\rangle$ as an initial state, let
us define $\bar{g}^{\rm ini}_k$ and $\bar{b}^{\rm ini}_k$ to be the initial
average amplitudes of the good and the bad states according to the oracle~$O_k$:
\begin{equation}
\bar{b}^{\rm ini}_k=B^{k-1}\,,
\end{equation}
\begin{align}
\bar{g}^{\rm ini}_k&=\frac{1}{N_k}\Big(A_1^{k-1}N_1+A_2^{k-1}(N_2-N_1)
+\dots+A_{k-1}^{k-1}(N_{k-1}-N_{k-2})+ B^{k-1}(N_k-N_{k-1}) \Big)\cr
&=\frac{\sum_{s=1}^{k-1}N_s h_s}{N_k}+B^{k-1} \;.
\end{align}

The initial separation, $\Delta^{\rm ini}_k$, between the good and the bad
averages is therefore
\begin{equation}\label{Delta0}
\Delta^{\rm ini}_k=\bar{g}^{\rm ini}_k-\bar{b}^{\rm ini}_k
=\frac{\sum_{s=1}^{k-1}N_s h_s}{N_k}\;.
\end{equation}
Observe that developing a new feature of height $h_k$ is equivalent to
increasing the initial separation $\Delta^{\rm ini}_k$ by $h_k$. The final
separation (after $t_k$ steps) between the good and bad averages is therefore
\begin{equation} \label{Delta_tk}
\Delta^{\rm fin}_k = \Delta^{\rm ini}_k + h_k 
= \frac{\sum_{s=1}^{k}N_s h_s}{N_k}\,.
\end{equation} 
Using (\ref{Delta0}) and (\ref{Delta_tk})
together with~(\ref{DeltaT}),
we therefore have
\begin{equation} \label{ForTk}
t_k=\frac{1}{\omega_k}\left(\arcsin\Big(\frac{\Delta^{\rm fin}_k}{\alpha_k}
             \sqrt{\frac{N_k}{2^aN}}\Big)
                -\arcsin\Big(\frac{\Delta^{\rm ini}_k}{\alpha_k}
             \sqrt{\frac{N_k}{2^aN}}\Big)\right)\,,
\end{equation}
where
\begin{equation} \label{omega_k}
\omega_k=\arccos\Big(1-\frac{2N_k}{2^{a}N}\Big)\,,
\end{equation}
and
\begin{eqnarray}  \label{eq:alpha_kSquared}
\alpha_k^2&=&(\bar{b}^{\rm ini}_k)^2+\frac{(\bar{g}^{\rm ini}_k)^2 N_k}{2^aN-N_k}\cr
&=&\frac{\Big(\sum_{s=1}^{k-1}N_s h_s\Big)^2
+N_k\Big(1-\sum_{s=1}^{k-1}(N_s-N_{s-1})C_{s,k-1}^2\Big)
}{N_k(2^aN-N_k)}\;.
\end{eqnarray}

To achieve a good fidelity between the state $|\Psi_{\tilde p}\rangle$ that we
actually prepare and our ``target'' state $|\Psi_{p}\rangle$, we want the
features $h_k$ to be as close as possible to the target values $\delta_k$
defined in Eq.~(\ref{deltas}). This motivates the
formulas~(\ref{eq:alpha_k}--\ref{eq:t_k}) for $t_k$ which are obtained {from}
the formulas~(\ref{Delta0}--\ref{eq:alpha_kSquared}) by (i) replacing the
features $h_k$ by the targets $\delta_k$, (ii) replacing the widths $N_k$ by
the measured values $\tilde N_k$, and (iii) by rounding to the nearest
integer.

\section{Fidelity analysis} \label{sec:fidelity}

In the above description of the algorithm, we have not specified how to choose
the parameters $\eta_c$, $\eta_g$ and $a$ as a function of
$\epsilon$ (or, alternatively, of the initially given parameters $\lambda$ and
$\eta$). The optimal choice for these parameters depends on the estimates
$\tilde n_1,\ldots,\tilde n_{1/\epsilon}$ obtained in the quantum counting
step. In this section we provide a rather generous 
worst case analysis which shows
that the choice
\begin{equation}   \label{eq:worstCase}
\eta_c={\epsilon^5}/{54} \;,\;\; \eta_g=0.99\epsilon^2
\;,\;\; a=\lceil \log_2\frac{\eta_g}{\eta_c}-3\rceil
\end{equation}
guarantees, with probability greater than $1-\nu$, the fidelity bound
\begin{equation}   \label{eq:realFidelityBound}
|\langle\Psi_{\tilde p}|\Psi_p\rangle| > 1-\lambda \;.
\end{equation} 
This bound is valid for arbitrary values of the $\tilde n_k$. In most
actual applications, much larger values of the accuracy parameters $\epsilon$,
$\eta_g$ and $\eta_c$ will be sufficient to guarantee this fidelity  bound.

We now show that Eq.~(\ref{eq:t_k}), for the times $t_k$  which we
motivated in the previous section, implies the fidelity
bound~(\ref{eq:realFidelityBound})  under the assumption
that the parameters $\eta_c$, $\eta_g$ and $a$ are chosen as in
Eq.~(\ref{eq:worstCase}). 

Our starting point will be two sets of expressions for the $t_k$, namely the
definition of the $t_k$, Eqs.~(\ref{eq:alpha_k}--\ref{eq:t_k}), in terms of
the measured values $\tilde N_k$ and the target values $\delta_k$, and
Eqs.~(\ref{Delta0}--\ref{eq:alpha_kSquared}) above in terms of the actual
values $N_k$ and $h_k$. In subsection~\ref{sec:hk-minus-deltak} we will derive
an upper bound on the error $|\delta_k-h_k|$.
This bound shows how accurate our algorithm is
in achieving the target height, $\delta_k$, for the features $h_k$.
The overall accuracy of our algorithm, however, also depends on
how accurate it is in achieving the correct width of the features.
This accuracy is determined by the
fraction of $x$ values for which $p'(x)\neq p''(x)$ (see Figure~\ref{figure1}).
In subsection~\ref{sec:exceptions} we obtain an upper bound on this fraction.
In subsection~\ref{sec:stage1Fidelity}, we derive the fidelity
bound~(\ref{eq:realFidelityBound}) and an upper bound 
on the probability that the algorithm fails 
due to a nonzero outcome of the measurement of the
auxiliary qubits. And finally, in
subsection~\ref{sec:stage2Fidelity}, we show that the
bound~(\ref{eq:tildePhiBound}) on the phases $\tilde\phi(x)$ implies the
overall fidelity bound~(\ref{eq:properFidelityBound}).
 
\subsection{Bound on $|\delta_k-h_k|$}   \label{sec:hk-minus-deltak}

It is convenient to perform the proof of the bound on 
$|\delta_k-h_k|$ in three steps. For this we note that 
$t_k$ depend on the values of $\tilde{\gamma}_k^{\rm ini}$, 
$\tilde{\gamma}_k^{\rm fin}$ and $\tilde{\omega}_k$.
In subsection~\ref{subsec:RangeForGammas} we determine 
the range of possible values for $\tilde{\gamma}_k^{\rm ini}$ and 
$\tilde{\gamma}_k^{\rm fin}$ that corresponds to the uncertainty
in the measured values of $\tilde{N}_k$. Similarly in 
subsection~\ref{subsec:RangeForOmegas} we determine the error range
for $\tilde{\omega}_k$, and finally, in subsection~\ref{subsec:ProofOfTheBound}
we complete the proof of the bound.

\subsubsection{Error range for $\tilde{\gamma}_k^{\rm ini}$ and 
$\tilde{\gamma}_k^{\rm fin}$} \label{subsec:RangeForGammas}

Equation~(\ref{ForTk}) provides an explicit expression for $t_k$ in terms of
$\{h_k\}$ and $\{N_k\}$:
\begin{equation}\label{t_kF1}
t_k=F(\{h_s\},\{N_s\})\,.
\end{equation}
In what follows it will be convenient to 
use auxiliary quantities $\{\tau_k\}$ defined as
\begin{equation}\label{tau_kF1}
\tau_k=F(\{\delta_s\},\{N_s\})\,.
\end{equation}
The meaning of $\tau_k$ becomes clear in comparison with (\ref{t_kF1}):
$\tau_k$ are the time intervals that correspond to the target features
$\{\delta_k\}$.  Unlike $\{t_k\}$, $\{\tau_k\}$ are not necessarily integers.

It will be convenient to rewrite the definition of $t_k$ given in
Eqs.~(\ref{eq:alpha_k}--\ref{eq:t_k}) in a slightly modified form:
\begin{equation}    
t_k = \Big\lfloor {1\over2} + 
{1\over\tilde\omega_k}
(\arcsin\tilde\gamma^{\rm fin}_k - \arcsin\tilde\gamma^{\rm ini}_k )
\Big\rfloor\,,
\end{equation}
where we use the following definitions:
\begin{equation} \label{TildeOmega_k}
\tilde{\omega}_k=\arccos\Big(1-\frac{2\tilde{N}_k}{2^aN}\Big)\,,
\end{equation}
\begin{equation}
\tilde{\gamma}^{\rm fin}_k=\frac{\tilde{\Delta}^{\rm fin}_k}{\tilde{\alpha}_k}
             \sqrt{\frac{\tilde{N}_k}{2^aN}}\,,\ \ \  
\tilde{\gamma}^{\rm ini}_k=\frac{\tilde{\Delta}^{\rm ini}_k}{\tilde{\alpha}_k}
             \sqrt{\frac{\tilde{N}_k}{2^aN}}\,,
\end{equation}
\begin{equation}
\tilde{\Delta}^{\rm fin}_k=
\frac{\sum_{s=1}^{k}\tilde{N}_s \delta_s}{\tilde{N}_k}\,,
\ \ \
\tilde{\Delta}^{\rm ini}_k
=\frac{\sum_{s=1}^{k-1}\tilde{N}_s \delta_s}{\tilde{N}_k}\,,
\end{equation}
and
\begin{equation}
\tilde{\alpha}_k^2
=\frac{\Big(\sum_{s=1}^{k-1}\tilde{N}_s \delta_s\Big)^2
+\tilde{N}_k\Big(1-\sum_{s=1}^{k-1}(\tilde{N}_s-\tilde{N}_{s-1})B_{s,k-1}^2\Big)
}{\tilde{N}_k(2^aN-\tilde{N}_k)}\;,
\end{equation}
where
\begin{equation}
B_{s,k}=\sum_{j=s}^{k}\delta_j\,.
\end{equation}
It is also convenient to define
\begin{equation}
\gamma^{\rm fin}_k=
\frac{\sum_{s=1}^{k}N_s\delta_s}{\alpha_k\sqrt{2^aN{N}_k}}
             \ \ \ {\rm and} \ \ \ 
\gamma^{\rm ini}_k=
\frac{\sum_{s=1}^{k-1} N_s\delta_s}{\alpha_k\sqrt{2^aN{N}_k}}\,.
\end{equation}
In this notation Eq.~(\ref{tau_kF1}) can be rewritten as
\begin{equation}
\omega_k\tau_k=\arcsin\gamma^{\rm fin}_k 
-\arcsin\gamma^{\rm ini}_k\,.
\end{equation}

Directly {from} the definitions we have
\begin{equation}
\Big(\tilde{\gamma}^{\rm fin}_k\Big)^2
=\frac{\Big(1-\frac{\tilde{N}_k}{2^aN}\Big)\Big(\sum_{s=1}^{k}\tilde{N}_s\delta_s\Big)^2}{\Big(\sum_{s=1}^{k-1}\tilde{N}_s\delta_s\Big)^2
+\tilde{N}_k\Big(1-\sum_{s=1}^{k-1}(\tilde{N}_s-\tilde{N}_{s-1})B_{s,k-1}^2\Big)
}\,.
\end{equation}
By direct calculation we get
\begin{eqnarray}
\sum_{s=1}^{k-1}(\tilde{N}_s-\tilde{N}_{s-1})B_{s,k-1}^2
&=&\sum_{s=1}^{k-1}\tilde{N}_s\Big(\delta_s^2
+2\delta_s\sum_{j=s+1}^k\delta_j\Big)
+\tilde{N}_{k-1}\delta_k^2\cr
&=&\sum_{s=1}^{k-1}N_s\Big(1\pm\eta_c\frac{N}{N_s}\Big)\Big(\delta_s^
+2\delta_s\sum_{j=s+1}^k\delta_j\Big)
+N_{k-1}\Big(1\pm\eta_c\frac{N}{N_{k-1}}\Big)\delta_k^2 \;. \cr 
&&
\end{eqnarray}
The $\pm$ notation is an abbreviation for a double inequality (see the
appendix). Since $N_s\geq(\eta_g-\eta_c)N$ we obtain
\begin{equation}
\sum_{s=1}^{k-1}(\tilde{N}_s-\tilde{N}_{s-1})B_{s,k-1}^2
=\Big(1\pm\frac{\eta_c}{\eta_g-\eta_c}\Big)
\sum_{s=1}^{k-1}(N_s-N_{s-1})B_{s,k-1}^2\,,
\end{equation}
and similarly
\begin{equation} 
\sum_{s=1}^{k}\tilde{N}_s\delta_s
=\Big(1\pm\frac{\eta_c}{\eta_g-\eta_c}\Big)\sum_{s=1}^k N_s\delta_s\,.
\end{equation}
Since $\eta_c/\eta_g<1/10$, we therefore have
\begin{equation}\label{ApproxGammaTk}
\tilde{\gamma}^{\rm fin}_k
=
\gamma^{\rm fin}_k
\frac{(1\pm\frac{\eta_c}{\eta_g-\eta_c})^{3/2}}{1\pm\frac{\eta_c}{\eta_g-\eta_c}}
=\gamma^{\rm fin}_k\Big(1\pm 10\frac{\eta_c}{\eta_g}\Big)\,,
\end{equation}
and similarly
\begin{equation}\label{ApproxGamma0}
\tilde{\gamma}^{\rm ini}_k
=
\gamma^{\rm ini}_k\Big(1\pm 10\frac{\eta_c}{\eta_g}\Big)\,.
\end{equation}
In the above formulas we have used the mean value theorem that states that for
any function $f$ that is continuous on the interval $|x|\leq c$, where $c$ is
some constant, and differentiable on $|x|<c$ we can write
\begin{equation}
f(x)=f(0)+xf'(x_0)\,,
\end{equation}
where $f'(x_0)$ denotes the derivative of $f$ at some point $|x_0|<|x|$. This
gives, for example, that for $|x|<1/10$
\begin{eqnarray} \label{MeanValueExamples}
\frac{1}{1\pm x}&=&1\pm 2x\,,\cr
(1\pm x)^{3/2}&=&1\pm 1.6 x\,.
\end{eqnarray}
The error bounds~(\ref{ApproxGammaTk}) 
and~(\ref{ApproxGamma0}) were obtained by simplifying the somewhat tighter
but unwieldy bounds using these methods. \\

\subsubsection{Error range for $\tilde{\omega}_k$} \label{subsec:RangeForOmegas}

The aim of this subsection is to determine the ratio
between $\tilde{\omega}_k$ and the true value $\omega_k$
given by Equations (\ref{TildeOmega_k}) and 
(\ref{omega_k}) respectively.
Using the mean value theorem we have, by definition,
\begin{eqnarray}
\tilde{\omega}_k&=& 
\arccos\Big(1-\frac{2N_k}{2^aN}(1\pm\frac{\eta_c}{\eta_g-\eta_c})\Big)\cr
&&\cr
&=&
\arccos\Big(1-\frac{2N_k}{2^aN}\Big)
-\frac{1}{\sqrt{1-[1-\frac{2N_k}{2^aN}(1+u)]^2}}
\Big(\pm\frac{2N_k}{2^aN}\frac{\eta_c}{\eta_g-\eta_c}\Big)\,,
\end{eqnarray}
where $u$ is a real number such that $|u|<\eta_c/\eta_g$. This implies
\begin{equation}
\tilde{\omega}_k=\omega_k\pm\sqrt{\frac{2N_k}{2^aN}}
\frac{1}{\sqrt{1+u}\sqrt{2-\frac{2N_k}{2^aN}(1+u)}}\frac{\eta_c}{\eta_g-\eta_c}\,.
\end{equation}
Since $\eta_c/\eta_g<1/10$ we have
\begin{equation}
\frac{\eta_c}{\eta_g-\eta_c}<\frac{\eta_c}{0.9\eta_g}\,,\ \ \ \ \
\sqrt{1+u}>\frac{1}{0.9\sqrt{2}}\,,\ \ \ \ \ 
\sqrt{2-\frac{2N_k}{2^aN}(1+u)}>1\,,
\end{equation}
and therefore
\begin{equation}
\tilde{\omega}_k=\omega_k\pm 2\frac{\eta_c}{\eta_g}\sqrt{\frac{N_k}{2^aN}}\,.
\end{equation}
It now remains to find a bound on $\sqrt{\frac{N_k}{2^aN}}$
that is linear in $\omega_k$.
We have, by definition,
\begin{equation}
\frac{2N_k}{2^aN}=1-\cos\omega_k=2\sin^2\frac{\omega_k}{2}\,.
\end{equation}
Since $x^2\geq\sin^2 x$ we
obtain
\begin{equation} \label{Inequality112}
\sqrt{\frac{N_k}{2^aN}}\leq\frac{\omega_k}{2}\,,
\end{equation}
and therefore
\begin{equation}\label{ApproxOmega}
\frac{\tilde{\omega}_k}{\omega_k}=1\pm \frac{\eta_c}{\eta_g}\,.
\end{equation}

\subsubsection{Proof of the bound}\label{subsec:ProofOfTheBound}

During the $k$th stage
our algorithm creates a feature of height
\begin{equation}
h_k=\alpha_k\sqrt{\frac{2^aN}{N_k}}
     \Big(\sin(\omega_k t_k-\xi_k)-\sin(-\xi_k)\Big)\,,
\end{equation}
where $\xi_k$ is some initial phase.
On the other hand, the target height $\delta_k$  is
\begin{equation}
\delta_k=\alpha_k\sqrt{\frac{2^aN}{N_k}} \,
     \Big(\sin(\omega_k\tau_k-\xi_k)-\sin(-\xi_k)\Big)\,.
\end{equation}
Hence
\begin{eqnarray} \label{HeightError}
|h_k-\delta_k|&=&
\alpha_k\sqrt{\frac{2^aN}{N_k}} \,
\Big|\sin(\omega_k t_k-\xi_k)-\sin(\omega_k\tau_k-\xi_k)\Big|\cr
&&\cr
&\leq&2\alpha_k\sqrt{\frac{2^aN}{N_k}} \,
             \Big|\sin\frac{\omega_k(t_k-\tau_k)}{2}\Big|\cr
&&\cr
&\leq&\alpha_k\sqrt{\frac{2^aN}{N_k}}\;\omega_k|t_k-\tau_k|\,.
\end{eqnarray}
Directly {from} the definition we obtain
\begin{equation}
\omega_k t_k=\frac{\omega_k}{\tilde{\omega}_k}
\Big(\arcsin\tilde{\gamma}^{\rm fin}_k
-\arcsin\tilde{\gamma}^{\rm ini}_k\Big)\pm\omega_k\,.
\end{equation}
Using Eqs.~(\ref{ApproxGammaTk}), 
(\ref{ApproxGamma0}), and~(\ref{ApproxOmega}),
using the fact that $10\eta_c/\eta_g\leq1/4$ and using 
the inequality~(\ref{ArcsinInequality})
we have
\begin{equation}
\omega_k t_k=\frac{1}{1\pm\eta_c/\eta_g}
\Big(\zeta_k\pm 4\sqrt{10\frac{\eta_c}{\eta_g}}\Big)\pm\omega_k\,.
\end{equation}
Now using~(\ref{MeanValueExamples}) and the fact that
$|\zeta_k|\leq\pi$ we derive {from} the above equation
\begin{equation}
\omega_k(t_k-\tau_k)=\pm 8\frac{\eta_c}{\eta_g}\pm\omega_k\,.
\end{equation}
Since $a\geq 3$ [see Eqs.(\ref{eq:worstCase}) and~(\ref{eq:epsilonBound})], we
have that $2N_k/(2^aN)\leq 1/4$, and we can therefore use the
inequality~(\ref{ArccosInequality}) to show that
\begin{equation}
\omega_k=\pm 2\sqrt{\frac{2N_k}{2^aN}}\,.
\end{equation}
Since $a\le\log_2(\eta_g/\eta_c)-3$, we obtain the bound
\begin{equation} \label{PhaseError}
\omega_k(t_k-\tau_k)=\pm 16\frac{\eta_c}{\eta_g}\,.
\end{equation}

The maximum possible value of the average
amplitude of "bad" states is
\begin{equation}
\max\bar{b}=\frac{1}{\sqrt{2^aN}}\,,
\end{equation}
and the maximum possible value of the average
amplitude of "good" states is
\begin{equation}
\max \bar{g}=\frac{1}{\sqrt{\eta N}}\,.
\end{equation}
One can therefore write
\begin{eqnarray}
\alpha^2_k&<&(\max\bar{b})^2+\frac{(\max\bar{g})^2N_k}{2^aN-N_k}\cr
&\leq&\frac{1}{2^aN}+\frac{1}{\eta(2^a-1)N}\cr
&\leq&\frac{1}{2^aN}+\frac{2}{\eta 2^a N}\cr
&<&\frac{4}{\eta 2^aN} \;.
\end{eqnarray}
Using this bound together with~(\ref{PhaseError})
we obtain for the error~(\ref{HeightError}):
\begin{equation}
|h_k-\delta_k|<\frac{32}{\sqrt{\eta N(1-\eta_c/\eta_g)}}
\cdot\frac{\eta_c}{\eta_g^{3/2}}\,.
\end{equation}
Substituting the parameters~(\ref{eq:worstCase}) into the right-hand side of
the above inequality one can show that
\begin{equation} \label{h_delta_bound}
|h_k-\delta_k|<\frac{\epsilon^2}{\sqrt{\eta N}}\,.
\end{equation}

\subsection{Number of exceptional values}   \label{sec:exceptions}

When describing our algorithm in section~\ref{sec:Algorithm} we have
introduced functions $p'$ and $p''$ to distinguish two different
approximations to the target function $p$. Namely, if $p'$ is an
approximation of $p$ which is defined by the oracles $o_k$, then 
$p''$ also takes into account the fact that we may not know the 
exact values of $n_k$. In our algorithm we therefore use $p''$
as our target function which coincides with $p'$ everywhere
apart for a small fraction of values of $x$ for which
$p'(x)\neq p''(x)$. In this section we obtain an upper
bound on this fraction which will then be used in 
the next section where we derive bounds on the fidelity
and the failure probability.

For all $j$, we have  $|n_j-\tilde n_j|\le\eta_cN$.
For $j<f_1$, we have $\tilde n_j< \eta_g N$, and hence 
\begin{equation}
n_j< (\eta_g+\eta_c) N \;.
\end{equation} 
We now consider, for each $k\in\{2,\ldots,T\}$, all values of $j$ such that 
$f_{k-1}\le j< f_k$. For these values of $j$, we have
\begin{equation}
\tilde n_j \le \tilde n_{f_{k-1}} \;.
\end{equation} 
Since $n_j\ge n_{f_{k-1}}$, we have
\begin{equation}
|\tilde n_j - \tilde n_{f_{k-1}}| \le 2\eta_cN \;, 
\end{equation} 
hence
\begin{equation}
|\tilde n_j - \tilde N_{k-1}| \le 2\eta_cN \;, 
\end{equation} 
and finally
\begin{equation}
|n_j -  N_{k-1}| \le 4\eta_cN \;.
\end{equation} 
Since $T\le1/\epsilon$, we find that $p''(x)\ne p'(x)$ 
for at most $\mu N$ values, where
\begin{equation} \label{mu}
\mu = 4\eta_c/\epsilon + (\eta_g+\eta_c)\,.
\end{equation}

\subsection{Fidelity bound and the failure probability}  \label{sec:stage1Fidelity}

For $k=T$, Eq.~(\ref{PsiKminusOne}) can be rewritten in the form
\begin{equation}
|\Psi^T\rangle
=\sum_{x=0}^{2^aN-1}\Big(B^T+\sum_{j=1}^T c_j(x)h_j\Big)
 |x\rangle\,,
\end{equation}
where
\begin{equation}
c_j(x)=\left\{\begin{array}{ll}
                 1& {\rm if}\ x<N_j\cr
                         0& {\rm otherwise}\,.
              \end{array}
           \right.
\end{equation}
Let us define
\begin{equation} \label{d}
d(x)=\sum_{j=1}^T c_j(x)(h_j-\delta_j)\,.
\end{equation}
Using this definition we have
\begin{eqnarray}
|\langle\Psi_p|\Psi^T\rangle|
&=& 
\left|\sum_{x=0}^{2^aN-1}\sqrt{p(x)}\,\Big(B^T+\sum_{j=1}^T c_j(x)h_j\Big)\right|\cr
&\geq& 
\left|\sum_{x=0}^{N-1}\sqrt{p(x)}\,\Big(B^T+\sum_{j=1}^T c_j(x)\,\delta_j\Big)\right|
-\left|\sum_{x=0}^{N-1}\sqrt{p(x)}\; d(x)\right|\,,
\end{eqnarray}
where we have used the fact that $p(x)=0$ for $x\geq N$.
Using~(\ref{h_delta_bound}), and since $T\leq 1/\epsilon$ 
we obtain
\begin{equation} \label{bound_on_d}
|d(x)|<\frac{\epsilon}{\sqrt{\eta N}}\;.
\end{equation}
Because $\sqrt{p(x)}\leq 1/\sqrt{\eta N}$, this implies
\begin{equation}
|\langle\Psi_p|\Psi^T\rangle|\geq
\left|\sum_{x=0}^{N-1}\sqrt{p(x)}\,\Big(B^T+\sum_{j=1}^T c_j(x)\,\delta_j\Big)\right|
-\frac{\epsilon}{\eta}\;.
\end{equation}
Rewriting Eq.~(\ref{eq:pDoublePrime}) in terms of the coefficients $c_j(x)$, 
\begin{equation} \label{Ppp_in_terms_of_c}
\sqrt{p''(x)}=\sum_{j=1}^T c_j(x)\,\delta_j\,,
\end{equation}
we can write
\begin{equation}
|\langle\Psi_p|\Psi^T\rangle|\geq
\sum_{x=0}^{N-1}\sqrt{p(x)}\sqrt{p''(x)}
-|B^T|\sqrt{\frac{N}{\eta}}
-\frac{\epsilon}{\eta}\;.
\end{equation}
For all $x$ with a possible exception of at most
$\mu N$ values $p''(x)=p'(x)$ (see Sec.~\ref{sec:exceptions}). Let $\set{S}_e$
be the set of exceptional values of $x$ for which
$p''(x)\neq p'(x)$. We have
\begin{equation}
|\langle\Psi_p|\Psi^T\rangle|\geq
\sum_{x=0}^{N-1}\sqrt{p(x)}\sqrt{p'(x)}
-\sum_{x\in\set{S}_e}\sqrt{p(x)}\Big|\sqrt{p'(x)}-\sqrt{p''(x)}\Big|
-|B^T|\sqrt{\frac{N}{\eta}}
-\frac{\epsilon}{\eta}\;.
\end{equation}
Since $\sqrt{p(x)}$, $\sqrt{p'(x)}$ and $\sqrt{p''(x)}$
are all bounded from above by $1/\sqrt{\eta N}$
we obtain
\begin{equation}
|\langle\Psi_p|\Psi^T\rangle|\geq
\sum_{x=0}^{N-1}\sqrt{p(x)}\sqrt{p'(x)}
-|B^T|\sqrt{\frac{N}{\eta}}
-\frac{\epsilon+\mu}{\eta}\;.
\end{equation}
By definition of $p'$
\begin{equation}
|\sqrt{p'(x)}-\sqrt{p(x)}|\leq \frac{\epsilon}{\sqrt{\eta N}}\;,
\end{equation}
and since $p$ is normalized we get
\begin{equation}   \label{fidelity_bound_almost}
|\langle\Psi_p|\Psi^T\rangle|\geq
1-|B^T|\sqrt{\frac{N}{\eta}}
-\frac{2\epsilon+\mu}{\eta}\;.
\end{equation}
In order to continue we need to calculate $|B^T|$.
This can be done by examining the normalization condition
$\langle\Psi^T|\Psi^T\rangle=1$. This condition reads
\begin{equation}
\sum_{x=0}^{2^aN-1}
\Big(
B^T+\sum_{j=1}^T c_j(x)\,h_j
\Big)^2=1\,.
\end{equation}
Using~(\ref{d}) and~(\ref{Ppp_in_terms_of_c}) this
can be rewritten as
\begin{equation}
\sum_{x=0}^{2^aN-1}
\Big(\sqrt{p''(x)}
+B^T+d(x)
\Big)^2=1\,,
\end{equation}
or
\begin{equation}  \label{JustBeforeQuadratic}
\sum_{x=0}^{2^aN-1}
\Big(\sqrt{p'(x)}
+B^T+d(x)
\Big)^2
-\Lambda
=1\,,
\end{equation}
where
\begin{eqnarray}
\Lambda&=&\sum_{x\in\set{S}_e}
\left(\Big(\sqrt{p'(x)}
+B^T+d(x)
\Big)^2
-\Big(\sqrt{p''(x)}
+B^T+d(x)
\Big)^2\right)\cr
&=&2 B^T\sum_{x\in\set{S}_e}\Big(\sqrt{p'(x)}-\sqrt{p''(x)}\Big)
+\sum_{x\in\set{S}_e}\Big(p'(x)-p''(x)\Big)\cr
& &+ 2\sum_{x\in\set{S}_e}\Big(\sqrt{p'(x)}-\sqrt{p''(x)}\Big)d(x) \,.
\end{eqnarray}
Let us define
\begin{equation}
e(x)=\sqrt{p'(x)}-\sqrt{p(x)}\,.
\end{equation}
Since $p$ is normalized, equation~(\ref{JustBeforeQuadratic})
gives a quadratic equation for $B^T$:
\begin{equation}
(B^T)^2+2U B^T +V =0\,,
\end{equation}
where
\begin{equation}
U=\frac{1}{2^a N}
\left(
\sum_{x=0}^{N-1}\Big(\sqrt{p'(x)}+d(x)\Big)
+\sum_{x\in\set{S}_e}(\sqrt{p''(x)}-\sqrt{p'(x)})
\right)\;,
\end{equation}
and
\begin{align}
V=&\frac{1}{2^a N}
\sum_{x=0}^{N-1}\left(2\sqrt{p(x)}\Big(d(x)+e(x)\Big)+
\Big(d(x)+e(x)\Big)^2
\right)\cr
&+\frac{1}{2^a N}\sum_{x\in\set{S}_e}
\left(
p''(x)-p'(x)+2d(x)\Big(\sqrt{p''(x)}-\sqrt{p'(x)}\Big)
\right)\,.
\end{align}
Since $\sqrt{p'(x)}$ and $\sqrt{p''(x)}$
are bounded from above by $1/\sqrt{\eta N}$
and since $\set{S}_e$ contains at most
$\mu N$ elements (see Sec.~\ref{sec:exceptions}), we obtain with the help
of~Eq.(\ref{bound_on_d})
\begin{equation}
|U|\leq\frac{1+\epsilon+\mu}{2^a\sqrt{\eta N}}\;.
\end{equation}
Similarly, since $|e(x)|\leq\epsilon/\sqrt{\eta N}$ we
have
\begin{equation}
|V|\leq\frac{6\epsilon+4\epsilon^2+\mu}{2^aN\eta}\;.
\end{equation} 
Since $\mu<\epsilon^2$ and $2^a>6/\epsilon^2$,
see Eqs.~(\ref{mu}) and~(\ref{eq:worstCase})
respectively, we obtain
\begin{equation}
|B^T|\leq |U|+\sqrt{U^2+|V|}\leq\frac{2\epsilon^2}{\sqrt{\eta N}}\;.
\end{equation}
Together with Eq.~(\ref{fidelity_bound_almost}) this gives
the lower bound on the fidelity,
\begin{equation}
|\langle\Psi_p|\Psi^T\rangle|\geq 1-\frac{3\epsilon}{\eta}\;,
\end{equation}
where we have observed that $\epsilon<1/3$ and used the 
bounds $2^a<7/\epsilon^3$,
$\mu<\epsilon^2$, which follow from our
settings given in Eq.~(\ref{eq:worstCase}).
The failure probability is 
\begin{equation} \label{TheFailureProbability}
p_{\rm fail}=(2^aN-1)N |B^T|^2<\frac{28\epsilon}{\eta}<10\lambda\,.
\end{equation}

\subsection{Introduction of phases and the fidelity bound}   \label{sec:stage2Fidelity}

We now show that the choice $\lambda'={\epsilon'}^2/8$ together with the
inequality~(\ref{eq:tildePhiBound}), i.e., $|\tilde\phi(x) - \phi(x)|
\le\epsilon'/2$, implies the overall fidelity
bound~(\ref{eq:properFidelityBound}). The proof is straightforward.
\begin{eqnarray}
|\langle\tilde\Psi|\Psi\rangle| 
&=& \Big| \sum_x \sqrt{p(x)\tilde p(x)} \,
           \exp[2\pi i(\phi(x)-\tilde \phi(x))] \Big| \cr
&\ge& \sum_x \sqrt{p(x)\tilde p(x)} \, 
      \cos[\phi(x)-\tilde \phi(x)]  \cr
&\ge& \sum_x \sqrt{p(x)\tilde p(x)} \, 
      \big( 1- [\phi(x)-\tilde \phi(x)]^2/2 \big) \cr
&\ge& \sum_x \sqrt{p(x)\tilde p(x)} \, 
      ( 1-\epsilon'^2/8 ) \cr
&=& |\langle\Psi_{\tilde p}|\Psi_p\rangle| \;
      ( 1-\lambda' ) \cr
&>& 1-\lambda-\lambda' \;.
\end{eqnarray}

\section{Resources} \label{sec:resources}

In this section we provide worst case upper bounds
on the resources required by the algorithm.
We distinguish between the resources that are needed for the
state preparation part of the algorithm 
(subsection~\ref{subsec:StatePrep})
and the resources that are needed by 
the quantum counting that precedes
the actual state preparation (subsection~\ref{subsec:QCounting}).

\subsection{Resources needed for state preparation} \label{subsec:StatePrep}

\subsubsection{Auxiliary qubits}

{From} our settings~(\ref{eq:worstCase}) we obtain
\begin{equation}
\frac{\eta_g}{\eta_c}<54/\epsilon^3\,.
\end{equation}
We thus obtain for the number of auxiliary qubits
\begin{equation}
a\le\log_2(\eta_g/\eta_c)-3<3+\log_2\epsilon^{-3}\,.
\end{equation}

\subsubsection{Oracle calls}

Here we give an upper bound on the time resources needed by the algorithm. The
construction of one feature requires at most
\begin{equation}
\max(t_k)\leq\frac{2\pi}{\omega_k}
\end{equation}
oracle calls. Using inequality~(\ref{Inequality112})
we can therefore write
\begin{equation}
\max t_k\leq \pi\sqrt{\frac{2^aN}{N_k}}\,.
\end{equation}
{From}~(\ref{eq:worstCase}) we have
\begin{equation}
\eta_g-\eta_c>\frac{8}{9}\epsilon^2\,.
\end{equation}
Since there are at most $1/\epsilon$ features and because
$N_k\leq(\eta_g-\eta_c)N$ and $2^a<8/\epsilon^3$ we therefore have that the
total number of oracle calls, $n_{\rm oracle}$, satisfies the bound
\begin{equation}
n_{\rm oracle}\le
\frac{\max t_k}{\epsilon} < 
\frac{3\pi}{\epsilon^3\sqrt{{\epsilon}}}\;.
\end{equation}

\subsection{Resources needed for counting} \label{subsec:QCounting}

\subsubsection{Counting accuracy}

Consider an oracle $O$ on the set of $2^aN$ possible values of $x$.
Using  standard techniques we can count the number $M$ of solutions
of $O$ within the absolute error $\Delta M$
\begin{equation}
\Delta M <\Big{(}\sqrt{2^aNM}+\frac{N}{2^{m-a+2}}\Big{)}2^{-m}\,,
\end{equation}
where $m$ is the number of auxiliary qubits needed by the standard
quantum counting routine~\cite{Nielsen}.

We want $\Delta M<\eta_cN$, where $\eta_c$ is the counting
accuracy introduced earlier. This connects the desired
counting accuracy $\eta_c$ with the number of auxiliary
qubits $m$,
\begin{equation}
\eta_c=\Big{(}\sqrt{\frac{2^aM}{N}}+{2^{a-2}\lambda}\Big{)}\lambda\,,
\end{equation}
where $\lambda=2^{-m}$.
Solving this equation for $\lambda$ in the case $\lambda>0$, we have
\begin{equation}
\lambda=2^{1-a/2}\Big{(}\sqrt{y+\eta_c}-\sqrt{y}\Big{)}\,,
\end{equation}
where $y=M/N$.  We see that, the bigger the value of $a$,
the bigger $m$ has to be in order to give the
required counting accuracy $\eta_c$.
We therefore set $a$ to the minimum, i.e., $a=1$ (this doubles the
range of $x$ values to
ensure reliable counting, see e.g.~\cite{Nielsen}).

It is easy to check that the dependence of $\lambda$
on $y$ is monotonic. As we
vary $M$ in the range 0 to $N-1$, the corresponding values of
$\lambda$ vary between the limits $\sqrt{2\eta_c}$ and
$\sqrt{2}(\sqrt{1+\eta_c}-1)>\eta_c/2 $. It follows that the required
number of auxiliary working qubits needed for counting with accuracy
$\eta_c$
is $m<\log\eta_c^{-1}$. Thus we choose $m=\log\eta_c^{-1}$.
This choice guarantees the required accuracy of counting irrespective of
the true value of $M$.

\subsubsection{Counting probability}

The above counting procedure does not output the correct result with
probability 1.  For the procedure to work correctly with probability $1-\nu$ we
have to increase the number of auxiliary qubits {from} $m$ to $a_c$ which is
given by
\begin{equation}
a_c=m+\log_2(2+\frac{1}{2\nu})=\log_2\frac{1+4\nu}{2\nu\eta_c}\,.
\end{equation}
The number, $N_{\rm count}$, 
of oracle calls that is required by 
the counting procedure is
\begin{equation}
N_{\rm count}=2^{a_c}-1 \,.
\end{equation}
Substituting $m=\log_2\eta_c^{-1}$ and using
Eq.~(\ref{eq:worstCase}) we obtain
\begin{equation}
a_c<\log_2\frac{27(1+4\nu)}{\nu\epsilon^5} \,.
\end{equation}
Since there are at most $1/\epsilon$ features the total number of oracle calls
needed by the counting stage of our algorithm is bounded as
\begin{equation}
N_{\rm count}^{\rm total}\leq\frac{N_{\rm count}}{\epsilon}
<\frac{27(1+4\nu)}{\nu\epsilon^6}\,.
\end{equation}

\section{Summary and conclusions} \label{sec:conclusions}

In conclusion, we have described a quantum algorithm to prepare an arbitrary
state of a quantum register of $\log_2N$ qubits, provided the state is initially
given in the form of a classical algorithm to compute the $N$ complex
amplitudes defining the state. For an important class of states, the algorithm
is efficient in the sense of requiring numbers of oracle calls and
additional gate operations that are polynomial in the number of qubits.
The following
table lists, for each stage of the algorithm, upper bounds on the number of
oracle calls and the number of auxiliary qubits needed. 

\begin{tabular}{lcc}     
$\;$ & oracle calls & auxiliary qubits \vspace{2mm} \\ 
counting & 
 $\frac{27(1+4\nu)}{\nu\epsilon^6}$ & 
 $\log_2\frac{27(1+4\nu)}{\nu\epsilon^5}$ \vspace{2mm} \\ 
preparing $|\Psi_{\tilde p}\rangle$ & 
 $\frac{3\pi}{\epsilon^3\sqrt{{\epsilon}}}$ &
 $3+3\log_2{1\over\epsilon}$ \vspace{2mm} \\
preparing $|\tilde\Psi\rangle$ & ${1\over\epsilon'}$ & 0 \vspace{2mm}
\end{tabular}

The bounds are not
tight and can be improved by a more detailed error analysis.  The total
number of quantum gate operations depends on the implementation of the
oracles. It is proportional to the number of oracle calls times a factor
polynomial in $\log_2 N$ if the functions $p(x)$ and $\phi(x)$ can be
efficiently computed classically.

Depending on the nature of the function $p(x)$ and the prior information about
$p(x)$, the algorithm we have described in this paper can be
optimized in a number of ways. For instance, the counting stage is the most
expensive in terms of both oracle calls and additional qubits. If for some
reason the numbers $n_k$ characterizing the oracles are known in advance, the
counting stage can be omitted, leading to considerable savings. Furthermore,
in this case the fidelity bound can be guaranteed with probability 1, i.e., we
can set $\nu=0$.

In some cases the algorithm can be simplified if, instead of using the oracles
defined in Eq.~(\ref{eq:ok}), one uses oracles that return the
$k$-th bit of the expression $\sqrt{p(x)/\eta N}$. The general conclusions of
the paper continue to hold for this variant of the algorithm, 
which we analyze in detail in~Ref.~\cite{SoklakovSchack}.

Finally, by using generalizations of Grover's algorithm in which the
oracles and the inversion about the mean introduce complex phase factors
\cite{Long1999,Zalka-9902} it is possible to reduce the number of auxiliary
qubits needed in the preparation stage of the algorithm. This leads to a
reduction in the number of required oracle calls, and could also be
important in implementations where the number of qubits is the main limiting
factor.

\section{Acknowledgments}

This work was supported in part by the European Union IST-FET project EDIQIP.

\appendix

\section{Appendix}

\subsection{Notation for double inequalities}

In this paper we have made a frequent use of the
following convention.
Let $a$, $b$ and $c$ be three numbers. The notation 
\begin{equation}
a = b\pm c
\end{equation}
is then understood to be equivalent to the double inequality
\begin{equation} \label{DoubleInequality}
b-c \leq a \leq b+c \;.
\end{equation}
Furthermore, let $g$, $e$ and $F$ be functions. The notation 
\begin{equation}
h(x) = \sum_{x\in{\cal I}} F\left(g(x)\pm e(x)\right) 
\end{equation}
is then equivalent to the statement that $h(x)$ can be written in the form
\begin{equation}
h(x) = \sum_{x\in{\cal I}} F\left(f(x)\right) \;,
\end{equation}
where $f(x)=g(x)\pm e(x)$ for all $x\in{\cal I}$.

\subsection{Trigonometric inequalities}
Here we prove the following
inequalities
\begin{equation}\label{ArcsinInequality}
|\arcsin(x+\nu)-\arcsin(x)|\leq 2\sqrt{|\nu|}\,,\ \ \ \ |\nu|\leq 1/4\,,
\end{equation}
and
\begin{equation} \label{ArccosInequality}
|\arccos(x+\nu)-\arccos(x)|\leq 2\sqrt{\nu}\,,\ \ \ \ |\nu|\leq 1/4\,.
\end{equation}
Consider the case $\nu\geq0$, which implies
that for any $x$
\begin{equation}
\arcsin(x+\nu)-\arcsin(x)\geq 0\,.
\end{equation}
By inspection of $\arcsin$-function we have that
for $0\leq\nu\leq 1/4$ the maximum value
of the difference $(\arcsin(x+\nu)-\arcsin(x))$
is achieved for $x=-1$:
\begin{equation} \label{maxArcsinDifference}
\max_x\Big(\arcsin(x+\nu)-\arcsin(x)\Big)=
\arcsin(\nu-1)-\arcsin(-1)\,.
\end{equation}
In the case of $y<z$ the following equality
holds~\cite{GradsteynRyzhik}
\begin{equation}
\arcsin z-\arcsin y=\arccos\Big(\sqrt{1-y^2}\sqrt{1-z^2}+yz\Big)\,.
\end{equation}
Applying this equality to the right hand side 
of~(\ref{maxArcsinDifference}) we obtain
\begin{equation} \label{AfterGradshteynRyzhik}
\arcsin(x+\nu)-\arcsin(x)
\leq \arccos(1-\nu)\,.
\end{equation}
Let us now look for a constant $c$
such that
\begin{equation} \label{ArccosOneMinusNu}
\arccos(1-\nu)\leq c\sqrt{\nu}\,. 
\end{equation}
Since $\arccos$ is a decreasing function
the above requirement is equivalent to
\begin{equation} \label{OneMinusNu}
1-\nu\geq\cos(c\sqrt{\nu})\,.
\end{equation}
According to the mean value theorem,
there exists $u<c\sqrt{\nu}$ such that
\begin{equation}
\cos(c\sqrt{\nu})=1-\frac{c^2\nu}{2}\cos(u)\,,
\end{equation}
and therefore the requirement~(\ref{OneMinusNu})
can be rewritten as
\begin{equation}
1\leq \frac{c^2}{2}\cos(u)\,.
\end{equation}
It is clear that this requirement is guaranteed
to be satisfied if we set $c=2$.
Indeed, $2\cos u> 1$ for any nonnegative
$u<\pi/3$ which includes all possible values
of $u$ that can correspond to $c=2$ and $\nu\leq 1/4$. 
Since $c=2$ guarantees that~(\ref{ArccosOneMinusNu})
is satisfied, we obtain from~(\ref{AfterGradshteynRyzhik})
\begin{equation}
\arcsin(x+\nu)-\arcsin(x)
\leq 2\sqrt{\nu}\,,\ \ \ \nu\geq 0.
\end{equation}
The case of negative $\nu$ can be treated in an analogous
fashion leading to the inequality
\begin{equation}
\arcsin(x)-\arcsin(x+\nu)
\leq 2\sqrt{|\nu|}\,,\ \ \ \nu\leq 0.
\end{equation}
The required inequality~(\ref{ArcsinInequality})
follows  trivially.
Moreover, since
\begin{equation}
\arcsin x +\arccos x=\pi/2,
\end{equation}
we also obtain (\ref{ArccosInequality}) as required.


\end{document}